# SATELLITE ATM NETWORK ARCHITECTURAL CONSIDERATIONS AND TCP/IP PERFORMANCE[1]

Sastri Kota
Lockheed Martin Telecommunications
1272 Borregas Avenue
Bldg B/551 O/GB - 70
Sunnyvale, CA 94089
Phone: (408)-543-3140, Fax: (408)-543-3104
Email: sastri.kota@lmco.com

Rohit Goyal, Raj Jain
Department of Computer Information Science
The Ohio State University
2015 Neil Ave, DL 395
Columbus, OH 43210-1277
Phone: 614-292-3989, Fax: 614-292-2911
Email: {goyal,jain}@cis.ohio-state.edu

## 1 INTRODUCTION

The rapid advances in ATM technology and Ka-Band satellite communications systems will lead to a vast array of opportunities for new value added services. Examples of such services include interactive as well as distribution services such as video conferencing, transmission of audio/video and high resolution image documents. Current trends in satellite communications exhibit an increased emphasis on new services as opposed to point-to-point data communications. The new services gaining momentum include mobile services, direct broadcast, private networks and high-speed hybrid networks in which services would be carried via integrated satellite-fiber networks. To fully realize these integrated systems, it is essential that advanced network architectures be developed that seamlessly interoperate with existing standards, interfaces and higher layer protocols.

With the deployment of ATM technology, there is a need to provide interconnection of geographically dispersed ATM networks. Although ATM technology has been developed to provide an end-to-end transparent service over terrestrial networks, satellite-ATM systems will play a significant role in achieving global connectivity and statistical multiplexing gains while maintaining Quality of Service (QoS) requirements. The ATM paradigm is aimed at supporting the diverse requirements of a variety of traffic sources, and providing flexible transport and switching services in an efficient and cost-effective manner.

The growing interest in Satellite ATM networking is due to the several advantages offered by satellite communications technology [3, 8]. These include, (a) wide geographic coverage including interconnection of "ATM islands", (b) multipoint to multipoint communications facilitated by the inherent broadcasting ability of satellites, (c) bandwidth on demand or Demand Assignment Multiple Access (DAMA) capabilities, and (d) an alternative to fiber optic networks for disaster recovery options.

However, satellite systems have several inherent constraints. The resources of the satellite communication network, especially the satellite and the earth station have a high cost and must be used efficiently. *A crucial issue is that of the high end-to-end propagation delay of satellite connections.* Apart from interoperability issues, several performance issues need to be addressed before a transport layer protocol like TCP can satisfactorily work over satellite-ATM networks for large delay-bandwidth networks. With an acknowledgment and timeout based congestion control mechanism (like TCP's), performance is inherently related to the delay-bandwidth product of the connection. As a result, the congestion control issues for broadband satellite networks are somewhat different from those of low latency terrestrial networks.

The performance optimization problem can be analyzed from two perspectives – *network architectures and end-system architectures.* The network can implement a variety of mechanisms to optimize resource utilization, fairness and higher layer throughput. For ATM, these include enhancements like feedback control, intelligent drop policies to improve utilization, per-VC buffer management to improve fairness, and even minimum throughput guarantees to the higher layers [9]. At the end system, the transport layer can implement various congestion avoidance and control policies to improve its performance and to protect against congestion collapse. Several transport layer congestion control mechanisms have been proposed and implemented. The mechanisms implemented in TCP are slow start and congestion avoidance [14], fast retransmit and recovery, and selective acknowledgments [6].

The organization of this paper is two-fold. In the first part, we outline the network based architectural issues to be addressed for and integrated Satellite-ATM network model. We then illustrate the structure of the TCP protocol stack, as an example of a popular end system protocol, over the ATM Unspecified Bit Rate (UBR) service category. We present simulation results for TCP performance and buffer requirements over the satellite-ATM-UBR service, and provide guidelines on improving TCP performance in such situations.

---

[1] Proceedings of the 3rd Ka Band Utilization Conference, Sorrento, Italy, September 15-18, 1997

## 2 THE SATELLITE ATM NETWORK MODEL

Figure 1 illustrates a satellite-ATM network model represented by a ground segment, a space segment, and a network control segment. The ground segment consists of ATM networks which may be further connected to other legacy networks. The network control center (NCC) performs various management and resource allocation functions for the satellite media. Inter-satellite crosslinks in the space segment provide seamless global connectivity via the satellite constellation. The network allows the transmission of ATM cells over satellite, multiplexes and demultiplexes ATM cell streams for uplinks, downlinks, and interfaces to interconnect ATM networks as well as legacy LANs. The satellite-ATM network consists of a satellite-ATM interface device connecting the ATM network to the satellite system. This interface is used for resource allocation, call control, error control, traffic control etc.

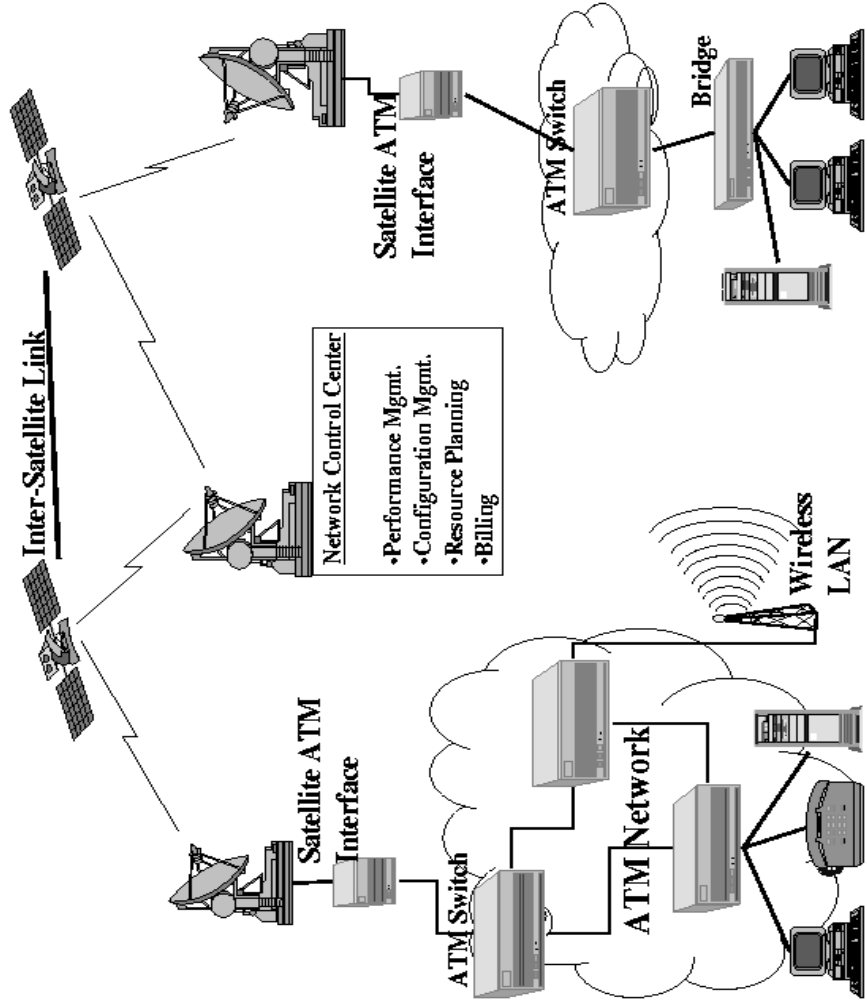

Figure 1: Satellite-ATM Network Model

Several issues need to be considered in designing various components of the above network architecture. The following subsections outline a few such issues, and provide recommendations on selecting the optimal implementation technology.

### 2.1 The ATM Quality of Service Model

ATM networks carry traffic from multiple service categories, and support Quality of Service (QoS) requirements for each service category. The ATM-Forum Traffic Management Specification 4.0 [1] defines five service categories for ATM networks. Each service category is defined using a traffic contract and a set of QoS parameters [1]. The *traffic contract* is a set of parameters that specify the characteristics of the source traffic. This defines the requirements for compliant cells of the connection. The traffic contract consists of:

*The source traffic descriptors.* These are used to specify the characteristics of the traffic from a source end system (SES). The Peak Cell Rate (PCR) specifies the maximum rate at which a source can send at any time. The Sustained Cell Rate (SCR) specifies the average rate maintained by the source. The Maximum Burst Size (MBS) specifies the maximum number of back to back cells at PCR that can be sent by the source without violation the SCR.

*The Cell Delay Variation Tolerance (CDVT) and Burst Tolerance (BT)* parameters are used to specify a tolerance for PCR and SCR respectively. The Generic Cell Rate Algorithm (GCRA) specified in [1] (a version of the leaky bucket algorithm) uses the PCR/SCR and the respective tolerance parameters to ensure that the incoming cells are compliant with the traffic contract. BT is calculated as BT = (MBS - 1)(1/SCR - 1/PCR).

The *QoS parameters* are negotiated by the source with the network, and are used to define the expected quality of service provided by the network. The parameters are:

*Maximum Cell Transfer Delay (Max CTD).* This is a measure of the maximum delay experienced by any cell of a connection within a switch. This delay consists of both a fixed cell processing delay and a variable queuing delay at a switch.

*Peak to Peak Cell Delay Variation (peak-to-peak CDV).* This is defined as the $100 \times (1-\alpha)$ quantile of the variable part of the delay. Peak-to-peak CDV places a bound on the variation in delay experienced by a cell in the switch.

*Cell Loss Ratio (CLR)* is defined as the ratio of the number of ATM cells that are discarded to the total number of cells transmitted.

For each service category, the network guarantees the negotiated QoS parameters if the end system complies with the negotiated traffic contract. For non-compliant traffic, the network need not maintain the QoS objective.

The *Constant Bit Rate (CBR)* class is defined for traffic that requires a constant amount of bandwidth, specified by PCR, to be continuously available. The network guarantees that all cells emitted by the source that conform to this PCR will be transferred by the network at PCR. The *real time Variable Bit Rate (VBR-rt)* class is characterized by PCR, SCR and MBS that controls the bursty nature of VBR traffic. The network attempts to deliver cells of these classes within fixed bounds of cell delay (max-CTD) and delay variation (peak-to-peak CDV). *Non-real-time VBR* sources are also specified by PCR, SCR and MBS, but are less sensitive to delay and delay variation than the real time sources. The network does not guarantee the CTD and CDV parameters for VBR-nrt.

The *Available Bit Rate (ABR)* service category is specified by a PCR as well as an MCR which is guaranteed by the network. The bandwidth allocated by the network to an ABR connection may vary during the life of a connection, but may not be less than MCR. ABR connections use a rate-based closed-loop feedback-control mechanism for congestion control. The network tries to maintain a low CLR by changing the allowed cell rates (ACR) at which a source can send.

The *Unspecified Bit Rate (UBR)* class is intended for best effort applications, and this category does not support any service guarantees. UBR has no built in congestion control mechanisms. The UBR service manages congestion by efficient buffer management policies in the switch.

These QoS parameters and service objectives have been specified by the ATM Forum. These values have to be re-evaluated for Satellite-ATM networks. The ITU-4B preliminary draft recommendations on transmission of Asynchronous Transfer Mode (ATM) Traffic via Satellite is in the process of development [5]. The maximum cell transfer delay of 400 ms for the ITU Class 1 stringent service needs to be reviewed to ensure that it properly accounts for the propagations delay for geosynchronous satellite networks. The peak-to-peak cell delay variation of 3 ms must also be carefully analyzed [4].

In the case of a satellite channel, bit errors are likely to occur in bursts due to the presence of encoding and decoding. The major performance objectives for B-ISDN links are specified in terms of acceptable ATM CLR. The stringent performance requirements are driven by the characteristics of optical fiber which can provide bit error rates (BERs) lower than $10^{-10}$. Moreover, the single bit containing the ATM header error correction (HEC) code is capable of correcting most errors encountered, given the random distribution of errors over fiber links. Satellite links that operate at high bandwidth (e.g., 155 Mbps) employ error correction schemes for providing acceptable link BER ($10^{-7}$ or better). The burst errors generated as a result of using these error correction schemes cannot be corrected by the ATM HEC since it is capable of correcting only single-bit errors. The ATM cell loss ratio over such links is therefore orders of magnitude higher than over links with random errors. Application of concatenated coding schemes with outer Reed-Solomon code and inner convolution code improves the error performance [2].

## 2.2 Media Access Protocols

Satellite bandwidth should be shared among user terminals fairly, flexibly and efficiently. Careful design of a media access control algorithm should be made, based on the the choice of the technologies in the

Table 1: Media Access Protocols

| Access protocol | Efficiency | Delay | Stability | Robustness | Complexity |
|---|---|---|---|---|---|
| S-ALOHA | 0.37 | Low | Low | High | Low |
| Tree CRA | 0.43-0.49 | Medium | Medium | Poor | Medium |
| DAMA (Reservation) | 0.6-0.8 | High | High | High | Medium |
| Hybrid (Reservation/Random) | 0.6-0.8 | Variable | Medium | High | Medium |

space and the ground segments. The key issues to be considered in selecting a media access protocol are:

*Efficiency or throughput.* This is the fraction of the time that useful traffic is carried over the multi-access channel.

*Access Delay.* This is the time between the arrival of a message and start of its successful transmission on the channel.

*Stability properties* relating to the possibility of undesirable long term congestion modes.

*Robustness* in the presence of channel errors and equipment failures.

*Implementation complexity* of the required hardware and software.

Table 1 provides a brief summary of a few candidate media access protocols. The Slotted ALOHA (S-ALOHA) scheme provides random access to the media and can result in instability. The Tree Contention Resolution Access (Tree CRA) protocol overcomes some problems of S-ALOHA, but is not very robust to failure. A bandwidth allocation scheme such as Demand Assignment Multiple Access (DAMA) can be employed to efficiently utilize the satellite link. DAMA allows the user to define the bandwidth that will actually be used, and thus improves network utilization. A hybrid scheme combines the use of initial random access of slots with reservation policies for following slots, and can provide low delay for short messages.

## 2.3 Traffic Management Issues

Traffic management tries to maximize traffic revenue subject to constraints of traffic contracts, QoS and "fairness". The traffic management problem is especially difficult during periods of heavy load particularly if traffic demands cannot be predicted in advance. For this reason, congestion control is an essential part of traffic management.

Congestion control is critical to both ATM and non-ATM networks. Several congestion control schemes provide feedback to the hosts to adjust their input rates to match the available link capacity. One way to classify congestion control schemes by the layer of the ISO/OSI reference model at which the scheme operates. For example, there are datalink, network and transport layer congestion control schemes. Typically, a combination of such schemes is used both by networks and end systems. The effectiveness of a scheme depends heavily upon factors like the severity, duration and location of the congestion [7].

# 3 TCP/IP TRAFFIC OVER A SATELLITE-ATM NETWORK

The ATM Unspecified Bit Rate (UBR) service category is expected to be used by a wide range of applications. Broadband switches should be able to multiplex thousands of transport connections that use UBR virtual circuits (VCs) for non-real time applications. On-board satellite switches, and switches at the earth stations fall into this category and are expected to multiplex a large number of non-real time transport connections over UBR virtual circuits. Figure 2 illustrates the protocol stack for Internet protocols over satellite-ATM. The satellite-ATM interface device separates the existing SONET and Physical Layer Convergence Protocol (PLCP) [3, 12]. Studies have shown that small switch buffer sizes result in very low TCP throughput over UBR [9]. It is also clear, that the buffer requirements increase with increasing delay-bandwidth product of the connections (provided the TCP window can fill up the pipe). However, the studies have not quantitatively analyzed the effect of buffer sizes on performance. *As a result, it is not clear how the increase in buffers affects throughput, and what buffer sizes provide the best cost-performance benefits for TCP/IP over UBR.* In this section, we present our simulation results to assess the buffer requirements for various satellite delay-bandwidth products for TCP/IP over UBR.

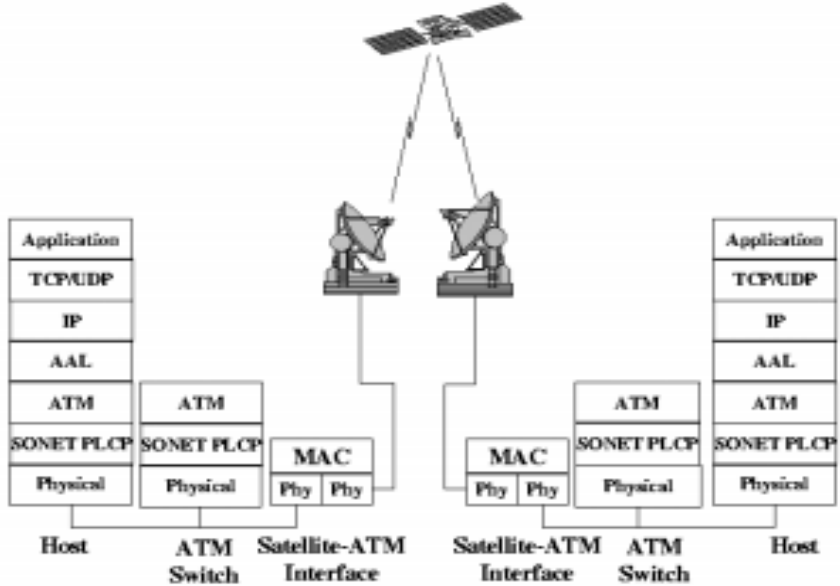

Figure 2: TCP over Satellite-ATM Protocol Stack

## 3.1 Performance Metrics

In our previous work [9, 10], we have studied TCP performance over the ATM-UBR service for terrestrial and satellite networks. In our studies, we have used an N-source symmetrical TCP configuration with unidirectional TCP sources. The performance of TCP over UBR is measured by the efficiency and fairness which are defined as follows:

$$\text{Efficiency} = (\text{Sum of TCP throughputs})/(\text{Maximum possible TCP throughput})$$

The TCP throughputs are measured at the destination TCP layers. Throughput is defined as the total number of bytes delivered to the destination application, divided by the total simulation time. The results are reported in Mbps.

The maximum possible TCP throughput is the throughput attainable by the TCP layer running over UBR on a 155.52 Mbps link. For 9180 bytes of data (TCP maximum segment size), the ATM layer receives 9180 bytes of data + 20 bytes of TCP header + 20 bytes of IP header + 8 bytes of LLC header + 8 bytes of AAL5 trailer. These are padded to produce 193 ATM cells. Thus, each TCP segment results in 10229 bytes at the ATM Layer. From this, the maximum possible throughput = 9180/10229 = 89.7% = 135 Mbps approximately on a 155.52 Mbps link (149.7 Mbps after SONET overhead).

$$\text{Fairness Index} = (\Sigma x_i)^2 / (N \times \Sigma x_i^2)$$

Where $x_i$ = throughput of the $i$th TCP source, and $N$ is the number of TCP sources. The fairness index metric applies well to our N-source symmetrical configuration.

## 3.2 Parameters

We study the effects of the following parameters:

> **Latency.** Our primary aim is to study the performance of large latency connections. The typical one-way latency from earth station to earth station for a single LEO (700 km altitude, 60 degree elevation angle) hop is about 5 ms [11]. The one-way latencies for multiple LEO hops can easily be up to 50 ms from earth station to earth station. GEO one-way latencies are typically 275 ms from earth station to earth station. We study these three latencies (5 ms, 50 ms, and 275 ms) with various number of sources and buffer sizes.

**Number of sources.** To ensure that the recommendations are scalable and general with respect to the number of connections, we will use configurations with 5, 15 and 50 TCP connections on a single bottleneck link. For single hop LEO configurations, we use 15, 50 and 100 sources.

**Buffer size.** This is the most important parameter of this study. The set of values chosen are $2^{-k} \times$ Round Trip Time (RTT), $k = -1..6$, (i.e., 2, 1, 0.5, 0.25, 0.125, 0.0625, 0.031, 0.016 multiples of the round trip delay-bandwidth product of the TCP connections.) We plot the buffer size against the achieved TCP throughput for different delay-bandwidth products and number of sources.

**Switch drop policy.** We use a per-VC buffer allocation policy called Selective Drop (see [9]) to fairly allocate switch buffers to the competing connections. This scheme uses per-VC accounting to maintain the current buffer utilization of each UBR VC. A fair allocation is calculated for each VC, and if the VC's buffer occupancy exceeds its fair allocation, its subsequent incoming packet is dropped. The scheme maintains a threshold R, as a fraction of the buffer capacity K. When the total buffer occupancy (X) exceeds R×K, new packets are dropped depending on the $VC_i$'s buffer occupancy ($Y_i$). In the selective drop scheme, a VC's packet is dropped if

$$(X > R) \text{ AND } (Y_i \times N_a/X > Z)$$

where $N_a$ is the number of active VCs (VCs with at least one cell the buffer) and Z is another threshold parameter ($0 < Z \leq 1$) used to scale the effective drop threshold. We choose the values of Z and R to be 0.8 and 0.9 respectively.

**End system policies.** We use an enhanced version of TCP called SACK TCP for this study. SACK TCP improves performance by using selective acknowledgments for retransmission [10].

## 3.3 Simulation Model

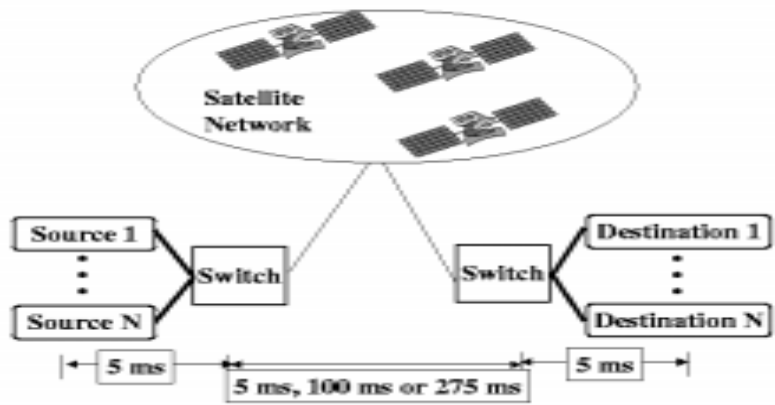

Figure 3: The N source TCP configuration

Figure 3 shows the basic network configuration that was simulated. In the figure, the switches represent the earth stations that connect to the satellite constellation. The entire satellite network is assumed to be a 155 Mbps ATM link without any on board processing or queuing. All processing and queuing are performed at the earth stations. All sources are identical, infinite and unidirectional TCP sources. Three different configurations are simulated that represent a single LEO hop, multiple LEO hops and a single GEO hop. The link delays between the switches and the end systems are 5 ms in all configurations. The inter-switch (earth station to earth station) propagation delays are 5 ms, 100 ms, and 275 ms for single hop LEO, multiple hop LEO and GEO configurations respectively. The maximum value of the TCP receiver window is 600000 bytes, 2500000 bytes and 8704000 bytes for single hop LEO, multiple hop LEO and GEO respectively. These window sizes are sufficient to fill the 155.52 Mbps links. The TCP maximum segment size is 9180 bytes. The duration of simulation is 100 seconds for multiple hop LEO and GEO and 20 secs for single hop LEO configuration. All link bandwidths are 155.52 Mbps, and peak cell rate at the ATM layer is 149.7 Mbps after the SONET overhead. The buffer sizes (in cells) used in the switch are the following:

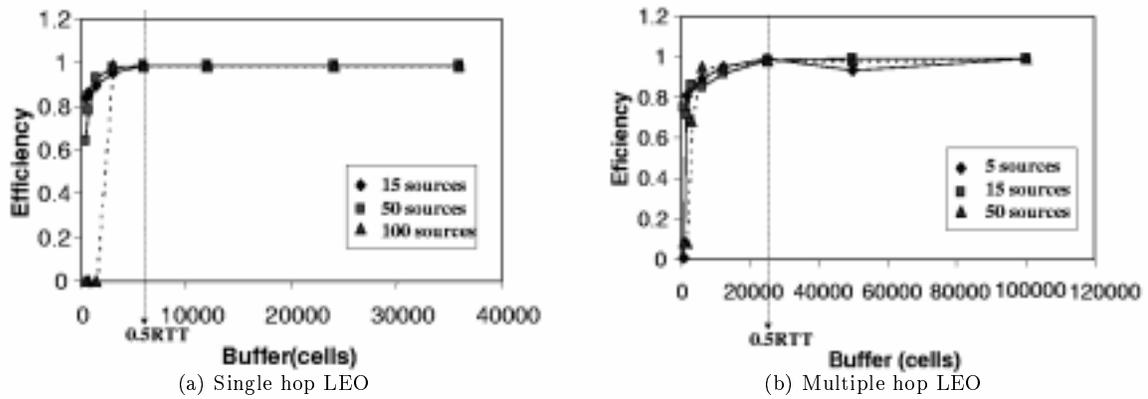

(a) Single hop LEO    (b) Multiple hop LEO

Figure 4: Buffer Requirements for LEO

Single LEO: 375, 750, 1500, 3 K, 6 K, 12 K (=1 RTT) , 24 K and 36 K.
Multiple LEO: 780, 1560, 3125, 6250, 12.5 K, 50 K (=1 RTT) , and 100 K.
GEO: 3375, 6750, 12500, 25 K, 50 K, 100 K, 200 K (=1 RTT) , and 400 K.

## 3.4 Simulation Results

Figures 4, and 5 show the resulting TCP efficiencies for the 3 different latencies. Each point in the figure shows the efficiency (total achieved TCP throughput divided by maximum possible throughput) against the buffer size used. Each figure plots a different latency, and each set of points (connected by a line) in a figure represents a particular value of N (the number of sources).

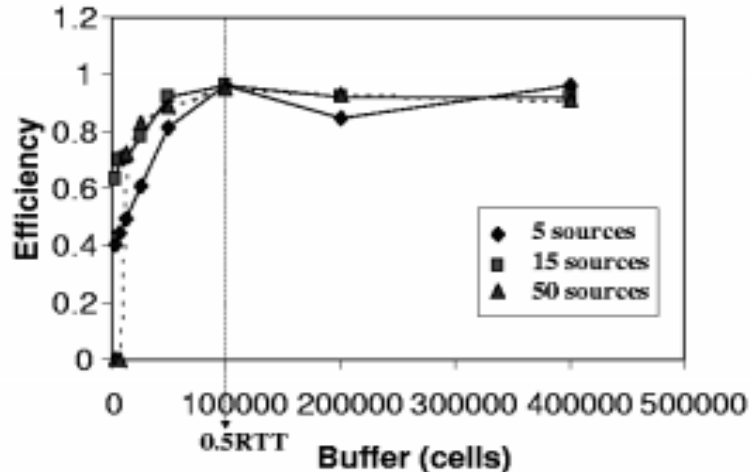

Figure 5: Buffer requirements for GEO

For very small buffer sizes, $(0.016 \times RTT, 0.031 \times RTT, 0.0625 \times RTT)$, the resulting TCP throughput is very low. In fact, for a large number of sources (N=50) , the throughput is sometimes close to zero. For moderate buffer sizes (less then 1 round trip delay-bandwidth), TCP throughput increases with increasing buffer sizes. TCP throughput asymptotically approaches the maximal value with further increase in buffer sizes. **TCP performance over UBR for sufficiently large buffer sizes is scalable with respect to the number of TCP sources.** The throughput is never 100%, but for buffers greater than $0.5 \times RTT$, the average TCP throughput is over 98% irrespective of the number of sources. Fairness is high for a large number of sources. This shows that TCP sources with a good per-VC buffer allocation policy like selective drop, can effectively share the link bandwidth.

## 4  Summary

In this paper, we have provided a summary of the design options in Satellite-ATM technology. A satellite ATM network consists of a space segment of satellites connected by inter-satellite crosslinks, and a ground segment of the various ATM networks. A satellite-ATM interface module connects the satellite network to the ATM networks and performs various call and control functions. A network control center performs various network management and resource allocation functions. Several issues such as the ATM service model, media access protocols, and traffic management issues must be considered when designing a satellite ATM network to effectively transport Internet traffic. We have presented the buffer requirements for TCP/IP traffic over ATM-UBR for satellite latencies. Our results are based on TCP with selective acknowledgments and a per-VC buffer management policy at the switches. A buffer size of about 0.5×RTT to 1×RTT is sufficient to provide over 98% throughput to infinite TCP traffic for long latency networks and a large number of sources. This buffer requirement is independent of the number of sources. The fairness is high for a large numbers of sources because of the per-VC buffer management performed at the switches and the nature of TCP traffic.

---

[2] All our papers and ATM Forum contributions are available from http://www.cis.ohio-state.edu/~jain